\documentclass[journal=jacsat,manuscript=article]{achemso}
\pdfoutput=1
\usepackage{amsmath}
\usepackage{amssymb}
\usepackage{hyperref}
\usepackage{textcomp}		
\usepackage{graphicx,caption}
\usepackage{xcolor}
\graphicspath{\Figures}
\usepackage{caption}
\usepackage{subcaption}
\usepackage{natbib}
\usepackage[utf8]{inputenc}
\usepackage{lscape}
\usepackage{pgfgantt}
\usepackage{geometry}
\usepackage{tikz}
\usetikzlibrary{shapes}
\usepackage{pgfplots}
\usepackage[version=3]{mhchem} 
\usepackage{comment}
\usepackage{lineno}
\usepackage{soul}

\newcommand{\circleK}{\raisebox{0.5pt}{\tikz{\node[draw,scale=0.4,circle,fill=black!20!black](){};}}}
\newcommand{\circleB}{\raisebox{0.5pt}{\tikz{\node[draw,scale=0.4,circle,fill=blue!20!blue](){};}}}
\newcommand{\squareK}{\raisebox{0pt}{\tikz{\node[draw,scale=0.4,regular polygon,regular polygon sides=4,fill=black!10!black](){};}}}
\newcommand{\squareB}{\raisebox{0pt}{\tikz{\node[draw,scale=0.4,regular polygon,regular polygon sides=4,fill=blue!10!blue](){};}}}
\newcommand{\squareG}{\raisebox{0pt}{\tikz{\node[draw,scale=0.4,regular polygon,regular polygon sides=4,fill=green!60!green](){};}}}
\newcommand{\redline}{\raisebox{2pt}{\tikz{\draw[-,red,solid,line width = 1.5pt](0,0) -- (5mm,0);}}}
\newcommand{\blueline}{\raisebox{2pt}{\tikz{\draw[-,blue,solid,line width = 1.5pt](0,0) -- (5mm,0);}}}
\newcommand{\blackline}{\raisebox{2pt}{\tikz{\draw[-,black,solid,line width = 1.5pt](0,0) -- (5mm,0);}}}
\newcommand{\bluelineDashed}{\raisebox{2pt}{\tikz{\draw[-,blue,dashed,line width = 1.5pt](0,0) -- (5mm,0);}}}

\newcommand{\bluelineS}{\raisebox{2pt}{\tikz{\draw[-,blue,solid,line width = 1.5pt](0,0.25mm) -- (1.5mm,0.25mm);}}}
\newcommand{\greenlineS}{\raisebox{2pt}{\tikz{\draw[-,green,solid,line width = 1.5pt](0,0.25mm) -- (1.5mm,0.25mm);}}}
\newcommand{\blacklineS}{\raisebox{2pt}{\tikz{\draw[-,black,solid,line width = 1.5pt](0,0) -- (1.5mm,0);}}}
\date{}

\author{Muhammad\,Rizwanur Rahman}
\affiliation{Department of Mechanical Engineering, Imperial College London, South Kensington Campus, London SW7 2AZ, United Kingdom}
\email{m.rahman20@imperial.ac.uk}
\author{Li Shen}
\affiliation{Department of Mechanical Engineering, Imperial College London, South Kensington Campus, London SW7 2AZ, United Kingdom}
\author{James P. Ewen}
\affiliation{Department of Mechanical Engineering, Imperial College London, South Kensington Campus, London SW7 2AZ, United Kingdom}
\author{Daniele Dini}
\affiliation{Department of Mechanical Engineering, Imperial College London, South Kensington Campus, London SW7 2AZ, United Kingdom}
\author{E. R. Smith}
\affiliation{Department of Mechanical and Aerospace Engineering, Brunel University London, Uxbridge UB8 3PH, United Kingdom}

\title{The Intrinsic Fragility of the Liquid-Vapor Interface: A Stress Network Perspective}
\begin{document}

\begin{abstract}
The evolution of the liquid-vapor interface of a Lennard-Jones fluid is examined with molecular dynamics simulations using the intrinsic sampling method. Results suggest clear damping of the intrinsic profiles with increasing temperature. Investigating the surface stress distribution, we identify a linear variation of the space-filling nature (fractal dimension) of the stress-clusters at the intrinsic surface with increasing surface tension, or equivalently, with decreasing temperature. A percolation analysis of these stress networks indicates that the stress field is more disjointed at higher temperatures. This leads to more fragile (or, poorly connected) interfaces which result in a reduction in surface tension.
\end{abstract}

\vspace{5mm}

\section{Introduction}

The intricate dynamics of the liquid-vapor interface have always attracted researchers from diverse research fields for its prevalence in myriad natural phenomena, such as bio-locomotion~\cite{hu2003hydrodynamics, bush2006walking,houghton2018vertically}, plastron respiration~\cite{flynn2008underwater}, and in numerous technological processes ranging from efficient oil recovery~\cite{nobakht2007effects} to organic electronics~\cite{forrest2007introduction} to capillarity driven thermal management~\cite{antao2016dynamic,rath2020core}.
To characterize the liquid-vapor interface, a number of approaches have been developed over the past century, from macro-scale treatment~\cite{de2013capillarity} to the sophisticated molecular dynamics (MD) simulations~\cite{ghoufi2016computer}, which have changed our perception of the interface quite dramatically~\cite{tarazona2012intrinsic}.

Among various available MD techniques, the mechanical route to the surface tension measurement has gained popularity for its relative simplicity and accuracy. This method replaces the scalar pressure with a second-rank pressure tensor~\cite{walton1983pressure} and thereby,
gives access to the atomic/molecular level detail of the interfacial region through the microscopic definition of the pressure. Following the \citeauthor{irving1950statistical}~(\citeyear{irving1950statistical}) definition, the diagonal elements of the pressure tensor can be expressed as:
\begin{equation}\label{Eq:IKpressure}
    \boldsymbol{P}(\boldsymbol{x},t) = \sum_{i=1}^{N} \frac{\boldsymbol {p}_i \boldsymbol{p}_i}{m_i} \delta(\boldsymbol{x}_i - \boldsymbol{x}) + \frac{1}{2} \sum_{i,j}^{N} \boldsymbol{f}_{ij}\boldsymbol{x}_{ij} O_{ij}\delta(\boldsymbol{x}_i -\boldsymbol{x})
\end{equation}
where, the momentum $\boldsymbol{p_i} = m_i (\boldsymbol{\dot{x}}_i -u)$, ${\boldsymbol{\dot{x}}}$ is the total particle velocity and $u$ is its streaming part, $\delta(\dots)$ is the Dirac delta function, $O_{ij}$ is the Irving Kirkwood operator~\cite{todd1995pressure}, $\boldsymbol{f}_{ij}$ is the force exerted by atom $j$ on atom $i$, and $\boldsymbol{x}_{ij} = \boldsymbol{x}_i -\boldsymbol{x}_j$ is the central distance between the two interacting atoms.
The average values of the off-diagonal elements in the pressure tensor are zero due to the axial symmetry about the normal direction and the translational symmetry about the plane parallel to the interface.
With the normal ($P_N$) and the tangential ($P_T$) components of the pressure, surface tension can be obtained from the Hulshof integral,~\cite{hulshof1901direct} as in Eq.~(\ref{Eq:surfTenMec}):
\begin{equation}\label{Eq:surfTenMec}
    \gamma  = \int_{-\infty}^{\infty} \left[ P_N (x) - P_T (x) \right] dx .
\end{equation}

Mechanical stability of the interface requires the normal component of the pressure tensor to be uniform and constant throughout the bulk phases and at the interface. On the contrary, the tangential component strongly depends on the position vector in the neighbourhood of the interface and only at a location far from the interface, this becomes uniform and equal to the normal component. Although Eq.\,(\ref{Eq:surfTenMec}) requires integration over the entire domain of the two-phase system, the term inside the integration soon becomes zero as one moves a few atomic diameters away from the interface. We will further elucidate this in the \nameref{sec:resultsndiscuss} section of this manuscript.

Attaining a molecular/atomic description of the interface, which is essential to understand the microscopic origin of surface tension~\cite{Antonin2011}, is challenging due to the inhomogeneous nature of the interface and the difficulties quantifying the surface forces.
The non-uniformity of the particle number density across the interface gives rise to mathematical complexity in uniquely identifying the system and its thermodynamic properties~\cite{malijevsky2012perspective}. 
In the past years, numerous efforts have been made, both experimentally~\cite{kuzmin1994influence,mitrinovic2000noncapillary} and by computer simulations~\cite{senapati2001computer,willard2010instantaneous, hofling2015enhanced,stephan2018vapor} to advance our understanding of the interface. A comprehensive review can be found in~\citet{ghoufi2016computer} and \citet{ghoufi2019calculation}.
Most MD simulation studies in the literature focused on the average profiles of the interface where the thermal fluctuation effects obscure the identification of the intrinsic behaviours~\cite{bresme2008molecular, braga2018pressure, willard2010instantaneous}.
The presence of thermal fluctuations, in the form of capillary waves, blurs the interfacial properties and makes it extremely difficult to extract the true nature of the interface~\cite{bresme2008molecular}.
Therefore, it is necessary to decouple the effects of capillary wave from the surface layer in order to investigate the intrinsic nature of the interface - an idea first introduced by \citet{buff1965interfacial} In their capillary wave theory, \citeauthor{buff1965interfacial} formalised the concept of the existence of an interface $\zeta (y,z,t)$ which acts as an instantaneous atomic border between the two phases, where the vector quantity, $y$ and $z$ are parallel to the interface. 
~\citet{sides1999capillary} exploited this idea of decoupling the capillary-wave broadening from the intrinsic interface and showed that the surface tension measured from Eq.\,(\ref{Eq:surfTenMec}) shows good agreement with that measured from the interface width.
However, to achieve a meaningful perspective of the intrinsic interface, adequate care must be taken in selecting the appropriate system size and the transverse resolution to avoid any blurring of the footprints of the intrinsic layer by the capillary wave fluctuations~\cite{chacon2006intrinsic}. 
Following its development, the intrinsic sampling method has been applied to understand the interface layer in a number of studies, i.e., to investigate the intrinsic gap between water-oil surface~\cite{bresme2008molecular} and to define the thickness of an adsorbed liquid film on a substrate~\cite{fernandez2011thickness}.

Given the strong dependency of the surface fluctuations on temperature, it is surprising that more attention has not been devoted to understand the effect of temperature variation on the intrinsic profiles of the liquid-vapor interface. Not only that temperature modifies the surface properties, often it becomes one of the key driving factors for a number of interfacial events, e.g., thermo-capillary flows.~\cite{jasnow1996coarse}
In the present study, we carefully examine the temperature effects on the intrinsic density and pressure profiles. Afterwards, 
we apply fractal analysis and the concept of percolating networks to analyse the temperature dependency of the spatial correlation of the interatomic interactions at the surface layer.

\section{Methodology}\label{sec:methodology}

MD simulations with the Flowmol MD code~\citep{smith2013coupling} were used to model the liquid-vapor interface of a Lennard-Jones (LJ)~\cite{jones1924determination} fluid at different temperatures.
The initial simulation domain was a cubic box of dimensions $L_x = 120.64$, $L_y =19.05$ and $L_z = 19.05$ in reduced LJ units, containing a total of $14,827$ particles. 
The middle 40\% of the box was initialized with LJ particles in liquid phase and the remaining was designated as vapor.
The initial state was created from a face-centred cubic (FCC) lattice with a starting density of $\rho=1$, then molecules were randomly deleted until the pre-set liquid ($\rho=0.5$) and vapor ($\rho=0.005$) densities were obtained.
Note, these initial units were somewhat arbitrary and used to set up the system only, a sufficient equilibration time was then allowed so the system tended to the expected coexistence (see Table S1 and Ref.~\citep{nistdata}). 
The equilibration phase was run in the canonical (NVT) ensemble using a No\'{s}e-Hoover thermostat,~\cite{nose1984molecular,hoover1985canonical} for 50,000 time steps with $\Delta T = 0.005$. A shifted LJ potential was used with periodic boundary conditions applied in all three Cartesian directions and the Verlet Leapfrog~\cite{verlet1967computer} integrator was used to integrate the equations of motion. 

For surface tension calculation, a small cutoff radius, $r_c$ results in significant deviation from experimental data for argon~\cite{nijmeijer1988molecular}, whereas, $r_c\geq 4$ is found to give better agreement~\cite{smith2016langevin}. Surface tension being one of our prime interests in this study, and as recommended by~\citet{ghoufi2016computer}, $r_c = 4.5$ was used to improve accuracy.
Further agreement with experimental data would require a more complex interaction potential, such as those which explicitly include three-body interactions.~\cite{ghoufi2016computer} The final state of the initial NVT calculations was taken as an initial condition for production runs in the microcanonical (NVE) ensemble. We used three independent simulations with different initial conditions to improve the statistics.

\begin{figure}
    \centering
    \includegraphics[width=0.9\linewidth]{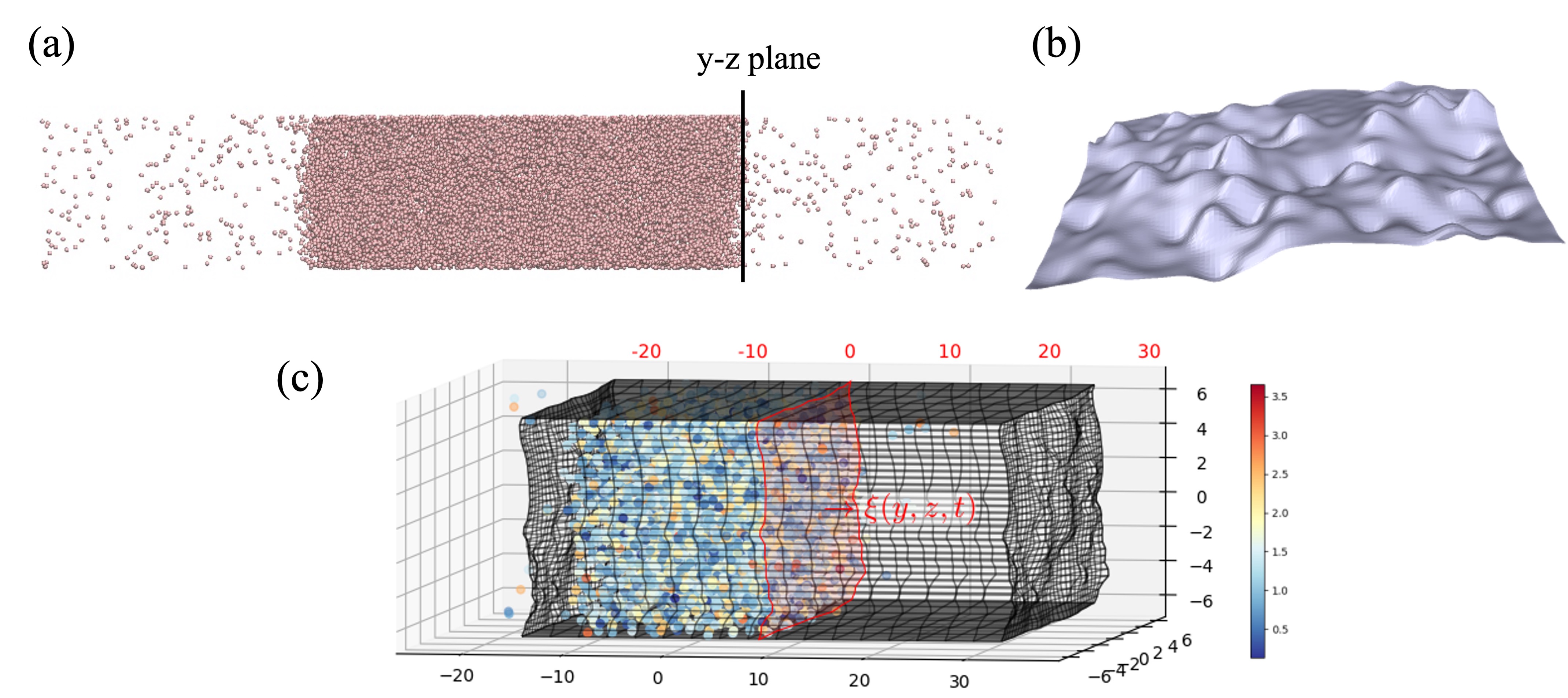}
     \caption{[Color online] (a) Liquid-vapor coexistence (b) fitted liquid-vapor surface function $\xi$ ($yz$ plane) (c) mapped grids at the interface, $x=0$ (mapped coordinates are in red at the top and un-mapped coordinates are in black at the bottom). The atoms are colored by their respective velocities, the bin-resolution is reduced here for visualization purpose. The mapping shifts every time the interface moves in the $x$ direction.}
    \label{fig:surf}
\end{figure}

The intrinsic interface was fitted to the outermost layer of the liquid by cluster analysis with the Stillinger cut-off length,~\cite{stillinger1963rigorous} $r_d = 1.5$, and with the required criterion of each atom having more than three neighbours.~\cite{smith2020hydrodynamics}
Instead of assuming an average interaction contour, the functional form of the interface, $\xi (y,z,t)$ was refitted at every time step using the intrinsic sampling method (ISM),~\cite{chacon2003intrinsic}  which approximates the liquid-vapor interface by means of a Fourier series representation as in Eq.\,(\ref{eq:fourier}).

\begin{equation}\label{eq:fourier}
    \xi (y,z,t) = \sum_{\textbf{k}<k_u} \hat{\xi_\textbf{k}} (t) ~\mbox{exp}(2\pi i \textbf{k} \cdot \boldsymbol{r}_\parallel)
\end{equation}
Here, $\boldsymbol{k}$ is the wave vector that corresponds to the periodic boundary conditions, i.e., $k = 2 \pi (n_y/L_y, n_z/L_z)$ with $n_y, n_z = 0, \pm 1, \pm 2, ... ,~k_u$ is the modulus of the wave vector, and $\boldsymbol{r}_\parallel$ denotes the parallel surface components in $y$ and $z$ directions.
The fitted coefficients, $\hat{\xi_\textbf{k}}$ are expressed as  time dependent functions because they are refitted to the surface each time the position of the surface atoms change (i.e. every time step). 
Fig.~\ref{fig:surf}\,(a) is representative of the coexisting system, where an intrinsic interface was fitted to the outermost liquid layer, the interface and the fluctuations are illustrated in Fig.~\ref{fig:surf}\,(b). 
All results presented hereafter correspond to the interface at the right side of the domain; $x>0$ corresponds to the vapor side and $x<0$ corresponds to the liquid side.

Once we have a mathematical form of the intrinsic interface, the density and pressure tensor can be obtained in a reference frame that moves with the surface, $\xi$ which has a different value at every point in $y$ and $z$ updated at every time step, $t$.
To obtain quantities that move with the interface, the \citet{irving1950statistical} definitions can be integrated over a volume where the surfaces in the $x$ direction follow the function given by Eq.\,(\ref{eq:fourier}), with uniform grids in $y$ and $z$ directions.
The density in a volume moving with the interface is then,
\begin{equation}
    \int_V \rho\, dV = \displaystyle\sum_{i=1}^N m_i \vartheta_i
\end{equation}
where $\vartheta_i$ is the product of Heaviside functions which is unity when an atom is inside a volume and zero if outside \cite{smith2020hydrodynamics}.
This can be expressed in terms of this product, $\vartheta_i = \Lambda_x(x_i) \Lambda_y(y_i) \Lambda_z(z_i) $, where, the difference between two Heaviside functions is known as a boxcar function and represented by $\Lambda$. This is formally obtained from integrating the Dirac delta function between finite limits but can be simply expressed in terms of an indicator function as follows,
\begin{linenomath}
    \begin{align}
        \Lambda_a (x) = 1_{a^- < x < a^+} (x) = \left\{
            \begin{array}{ll}
                1 & \quad a^- < x < a^+ \\
                0 & \quad \mathrm{otherwise},
            \end{array}
        \right.
        \label{Indicator_fn_boxcar} 
    \end{align}
\end{linenomath}     
which checks whether $x$ is between the two limits, $a^-$ and $a^+$.
In the $y$ and $z$ direction, these indicator functions $\Lambda_y(y_i)$ and $\Lambda_z(z_i)$ check whether the atom positions $\{y_i, z_i\}$ are between the spatial extents of a surface bounded by $y^-$ and $y^+$ and $z^-$ and $z^+$ respectively.

For the $x$ direction, the position of the interface is included in these limits, so for a box centred on $x$ with size $\Delta x$ the limits are $x^{\pm} = x \pm \Delta x/2 + \xi(y_i, z_i)$.
The easiest way to obtain the density relative to the intrinsic surface is to map the atomic positions based on the intrinsic surface at the same $y$ and $z$ location, $x_i^{'} = x_i - \xi(y_i, z_i)$ and then simply bin as if using a uniform grid, i.e., ${\tt{bin} = round}(x^{'}/ \Delta x)$.
The grid is uniform in the other two directions so the binning in $y$ and $z$ is unchanged, see fig.~\ref{fig:surf}\,(c). 
Quantities such as momentum, temperature and pressure inside a volume \cite{cormier2001stress} can also be obtained using the same mapping approach, where the latter requires mapping of the line of interaction between atoms to get the configurational term \cite{smith2020hydrodynamics}.
However, only the pressure tensor obtained from surface fluxes, taken here over the surfaces of a binning volume, can be shown to satisfy the mechanical equilibrium condition, i.e., the normal pressure is constant when moving through the interface~\cite{smith2021importance}.
This form of pressure also includes a term for the movement of the intrinsic surface itself, which must be accounted for to ensure balance of the coarse-grained equations.
The surfaces of a volume are flat in the $y$ and $z$ directions and follow the interface in the $x$ direction. 
The surface pressure can be written as the sum of three contributions,
\begin{equation}
\boldsymbol{P}_{\alpha}^\text{Total} = \boldsymbol{P}_{\alpha}^\text{kin.} + \boldsymbol{P}_{\alpha}^\text{SM.} + \boldsymbol{P}_{\alpha}^\text{config.}
\end{equation}
where $\alpha \in \{x,y,z\}$ denotes the three directions where each is a vector $\boldsymbol{P}_{\alpha}=[P_{x \alpha}, P_{y \alpha}, P_{z \alpha} ]^T$, with the kinetic contribution, $\boldsymbol{P}^\text{kin.}_{\alpha}$, which comes from the momentum transport of atoms crossing the interface~\cite{berry1971molecular}, the configurational contribution, $\boldsymbol{P}^\text{config.}_{\alpha}$ that arises from the atomic interaction~\cite{walton1983pressure},  and $\boldsymbol{P}^\text{SM.}_{\alpha}$ $-$ a term due to the surface movement in time.

Introducing a surface normal vector, which for the flat surfaces in $y$ and $z$ directions are simply the standard basis unit vectors, $\boldsymbol{n}_y=\boldsymbol{e}_y$ and $\boldsymbol{n}_z=\boldsymbol{e}_z$, while for the intrinsic surface, it is given by $ {\boldsymbol{n}_{x}}=\boldsymbol{\nabla}_{s}\left(\xi-x_{s}\right)/||\boldsymbol{\nabla}_{s}\left(\xi-x_{s}\right)||$ where the subscript, $s$, denotes the derivative taken at the point of crossing $x_s$.
By assuming the convective term to be zero, $\rho \boldsymbol{u} \boldsymbol{u} = 0$, the three pressure contributions discussed above can be obtained in a molecular simulation as follows,
\begin{linenomath}
\begin{align}
\int_{t_1}^{t_2}  \boldsymbol{P}^\text{kin.}_{\alpha}  dt
 = & \frac{1}{\Delta S_{\alpha}}  \displaystyle\sum_{i=1}^N  m_i \boldsymbol{\dot{r}}_i \boldsymbol{r}_{i_{12}} \cdot 
 \frac{ {\boldsymbol{n}_{\alpha}} }{|\boldsymbol{r}_{12} \cdot {\boldsymbol{n}}_{\alpha} |} dS_{\alpha}
\nonumber  \\ 
\int_{t_1}^{t_2}  \boldsymbol{P}^\text{SM.}_{\alpha}  dt
 = &  \frac{1}{\Delta S_{\alpha}} \displaystyle\sum_{i=1}^N  m_i \boldsymbol{\dot{r}}_i  \vartheta_t 
\nonumber  \\ 
\boldsymbol{P}_{\alpha}^\text{config.}= & \frac{1}{2 \Delta S_{\alpha}}\sum_{i,j}^{N} \boldsymbol{f}_{ij}  \boldsymbol{{r}}_{ij} \cdot
\frac{ {\boldsymbol{n}_{\alpha}} }{|\boldsymbol{r}_{ij} \cdot {\boldsymbol{n}_{\alpha}}|}dS_{\alpha}
\label{P_equation}
 \end{align}
 \end{linenomath} 
where $\Delta S_{\alpha}$ is the surface area, $\boldsymbol{r_{i_{12}}} = \boldsymbol{r}_i(t_2) - \boldsymbol{r}_i(t_1)$ denotes the vector for the movement of an atom from time $t_1$ to $t_2$ while $\boldsymbol{r}_{ij}$ is the separation vector.
The $\vartheta_t$ term captures the movement of the surface by counting all atoms which enter or leave a volume as the intrinsic interface itself moves in time. Defined as $\vartheta_t = \Lambda_{\xi} (x_i)\, \Lambda_{y} (y_i)\, \Lambda_{z} (z_i)$, the surface evolution from the start of a timestep $\xi^- = \xi(t)$ to the end $\xi^+=\xi(t+\Delta t)$ is multiplied by an indicator function, i.e.,  Eq.\,(\ref{Indicator_fn_boxcar}).

The $dS_{\alpha}$ term in Eq.(\ref{P_equation}) is the derivative of the $\vartheta$ function, with respect to $\alpha=\{x,y,z\}$, and is only non-zero if the separation vectors $\boldsymbol{r}_{i_{12}}$ or $\boldsymbol{r}_{ij}$ are crossing the surface of the volume in question.  Without loss of generality, we consider the expression for the surface and the separation vector, $\boldsymbol{r}_{ij}$ in the $x$ direction as 
\begin{displaymath}
    dS_{x} = \displaystyle\sum_{k=1}^{N_{\text{roots}}}
    \Lambda_{\lambda} (\lambda_k)\, 
    \Lambda_{y}(y_{\lambda}(\lambda_k))\, 
    \Lambda_{z}(z_{\lambda}(\lambda_k)) 
    \label{surface_crossing}
\end{displaymath}
where $\lambda$ parameterises the line between atoms $\boldsymbol{r}_{\lambda} = \boldsymbol{r}_i + \lambda \boldsymbol{r}_{ij}$ with $\lambda_k$ the value at a point of crossing on a surface.
The function $\Lambda_{\lambda}$ therefore checks if a $\lambda_k$ value is between $\boldsymbol{r}_i$ ($\lambda^- = 0$) and $\boldsymbol{r}_j$ ($\lambda^+ = 1$), with the remaining functions checking if the point of crossing $\{y_{\lambda}(\lambda_k), z_{\lambda}(\lambda_k) \}$ is on the volume surface between $y^-$ and $y^+$ and $z^-$ and $z^+$.
The form of $dS_y$ and $dS_z$ are similar and atomic motions are parameterised in the same way with $\boldsymbol{r}_{\lambda} = \boldsymbol{r}_1 + \lambda \boldsymbol{r}_{i_{12}}$. 

The calculation of the pressure tensor, therefore, becomes a ray-tracing problem, namely getting all the intersections of the surfaces of a volume due to interatomic interactions and atomic crossings.
In order to accelerate the process of getting intersections on an intrinsic surface, for each binning volume, the Fourier function of Eq.~(\ref{eq:fourier}) is converted to a set of bilinear patches of the form
\begin{displaymath}
\xi^{BL} (y,z) = a_o + a_y y + a_z z + a_{yz} yz
\end{displaymath}
The intersection of a line and the bilinear patch is a local operation, which is much quicker than the root finding process on a full Fourier surface of Eq.\,(\ref{eq:fourier}).
The procedure to fit the intrinsic interface and to choose the number of bilinear patches, as well as to calculate the intersect are described in previous work ~\citep{smith2020hydrodynamics, smith2021importance}.

\section{Results and Discussions} \label{sec:resultsndiscuss}
\subsection{Temperature effects on density profiles}

\begin{figure}
    \centering
    \includegraphics[width=0.80\linewidth]{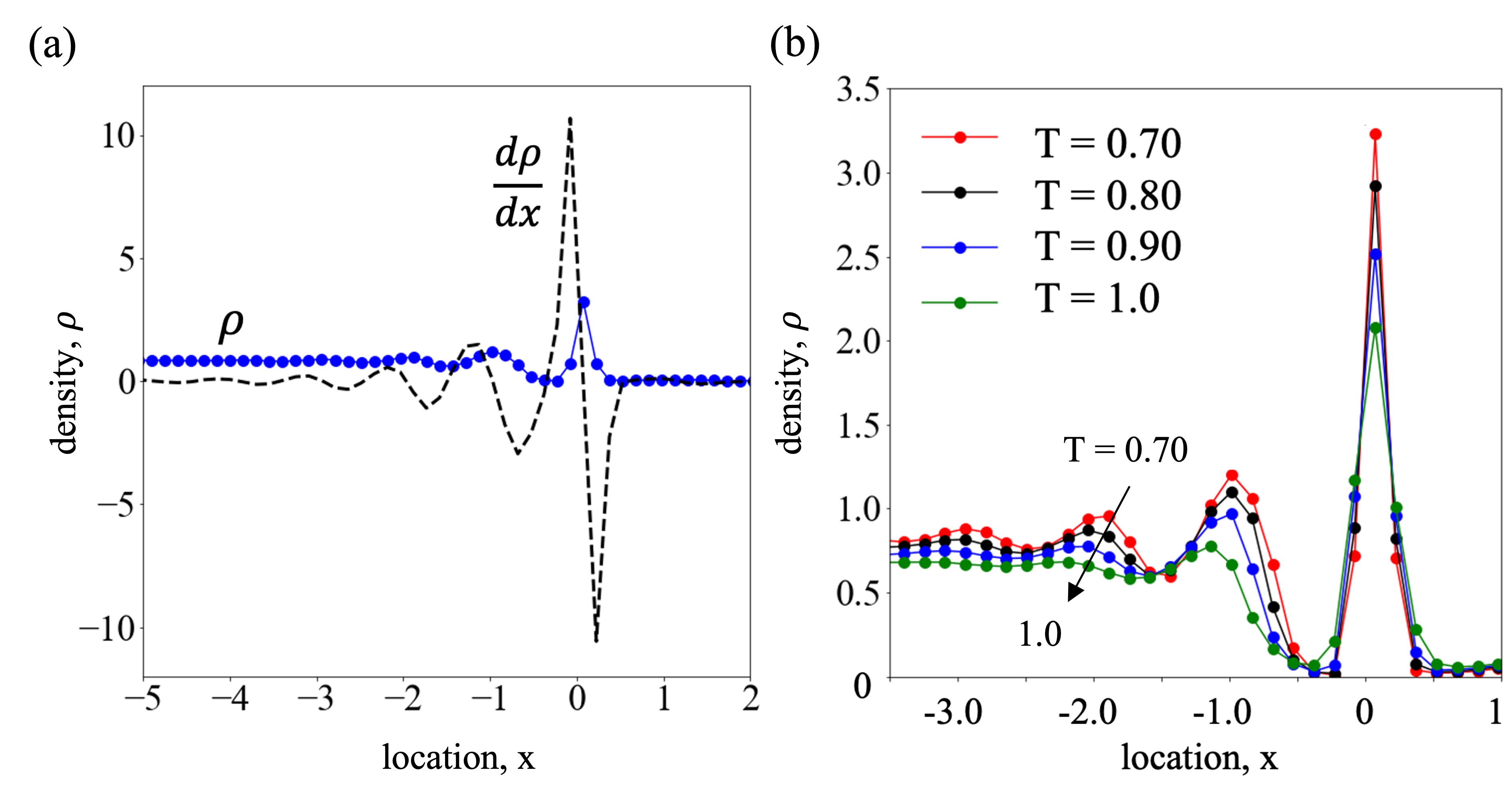}
    \caption{[Color online] (a) Density profile (blue circles) and derivative of density (black dashed line) for $T=0.70$ shows oscillations near the liquid-vapor interface (b) density profiles for a range of temperatures, oscillations are damped as temperature increases.}
    \label{fig:densityOsc}
\end{figure}

The intrinsic density profile at temperature $T=0.7$ is shown in Fig.~\ref{fig:densityOsc}\,(a). Considerable oscillations are evident near the intrinsic surface layer extending at least five atomic diameters into the liquid phase (density is averaged over time, as well as in the $y$ and $z$ spatial directions).
Although such layering is a universal behaviour of the free surface~\cite{chacon2001layering}, these oscillations are often smoothed out for an LJ fluid if the reference frame is static. On the other hand, a moving frame of reference makes the layering evident. This layering directly determines the stress that would be measured, both on the interface itself and in the bulk fluid. In the next sections, we focus on the effects of such layering on the surface stress and thereby, on the surface tension.

As seen from Fig.~\ref{fig:densityOsc}\,(a), the amplitude of the oscillation increases as the surface is approached from the liquid side, the highest peak is attained at the intrinsic surface and quickly dampens within a distance of less than half an atomic diameter into the vapor side. The first derivative of the density profile, $d\rho/dx$, further illustrates these fluctuations and the zero-crossing highlights the position of the intrinsic layer. Such derivatives are also instructive of the interfacial widths, readers are referred to~\citeauthor{sides1999capillary}\cite{sides1999capillary} for further details.
Fig.~\ref{fig:densityOsc}\,(b), where the intrinsic density profiles are plotted for a range of temperatures, shows increased damping at higher temperatures. Evidently, the fluctuations only exist at the liquid side, and dampens quite abruptly in the vapor side right after the interface ($x>0$), despite the existence of a small peak due to the adsorbed layer.

\begin{figure}
    \centering
    \includegraphics[width=0.80\linewidth]{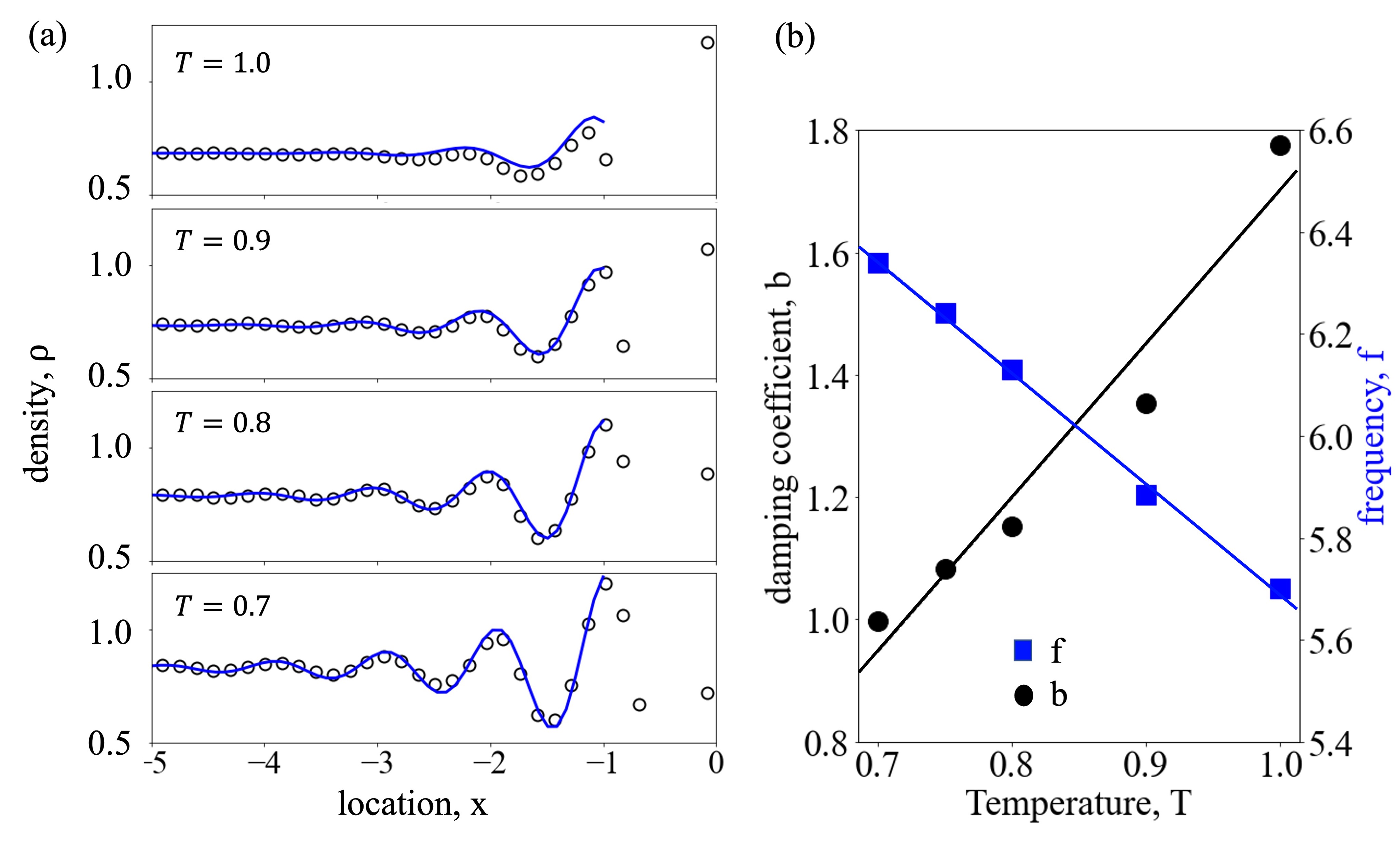}
    \caption{[Color online] (a) Oscillation of the intrinsic density profiles:  circles denote results from MD simulations, solid lines are exponential fitting, as in Eq.\,(\ref{eq:dampen}), to the density profiles at different temperatures, (b) the damping coefficient (left $y$ axis,\, \protect\circleK) increases with temperature, whereas, the frequency of oscillation (right $y$ axis,\,\protect\squareB) decreases. The solid lines are linear fits to the data.}
    \label{fig:osc_fitting}
\end{figure}

The oscillations of the density profile at the liquid side for $x \in (-\infty, -1]$ can be approximated, as illustrated in Fig.~\ref{fig:osc_fitting}\,(a), by an exponential decay function of the form:

\begin{equation}\label{eq:dampen}
\rho (x) = \rho_\mathrm{bulk} + e^ {bx} \cos (\mathrm{f}\, x)
\end{equation}where, $\rho (x)$ denotes the local intrinsic density, $\rho_\mathrm{bulk}$ is the liquid density of the bulk phase far away from the interface, $b$ is the damping coefficient, and $\mathrm{f}$ is the frequency of oscillation.
Note that the negative sign of the exponent, i.e., the damping coefficient in Eq.\,(\ref{eq:dampen}) is discarded as the location, $x$ in the liquid side is represented with negative numbers. The damping coefficient linearly increases, whereas the frequency of oscillation decreases as temperature is increased from $T=0.70$ to $1.0$, see Fig.~\ref{fig:osc_fitting}\,(b). 
Both the damping coefficient, $b$ and the frequency of the oscillation, $f$ show linear fits with temperature within the considered range, i.e., $b = 2.51\,T - 0.81$ with $R^2 = 0.95$ and $\mathrm{f} = -0.7\pi\,T + 2.5\pi$ with $R^2 = 0.99$. 
Thus the intrinsic density in Eq.\,(\ref{eq:dampen}) can be expressed as a function of location, bulk density of the liquid and the temperature, i.e.,
\begin{displaymath}
\rho(x) = \rho_\mathrm{bulk} + e^{(2.51T-0.81)x} \mbox{cos}\left[\pi(2.5-0.7T) x\right] ~\mathrm{for}~ x \in (-\infty, -1]
\end{displaymath}
\noindent Whilst a liquid-vapor interface is not directly comparable to a solid-liquid interface, one should obtain similar damping for the same bulk liquid in contact with a solid wall~\cite{evans1993asymptotic}.

\subsection{Temperature effects on pressure profiles}
\begin{figure}
    \centering
    \includegraphics[width=0.98\linewidth]{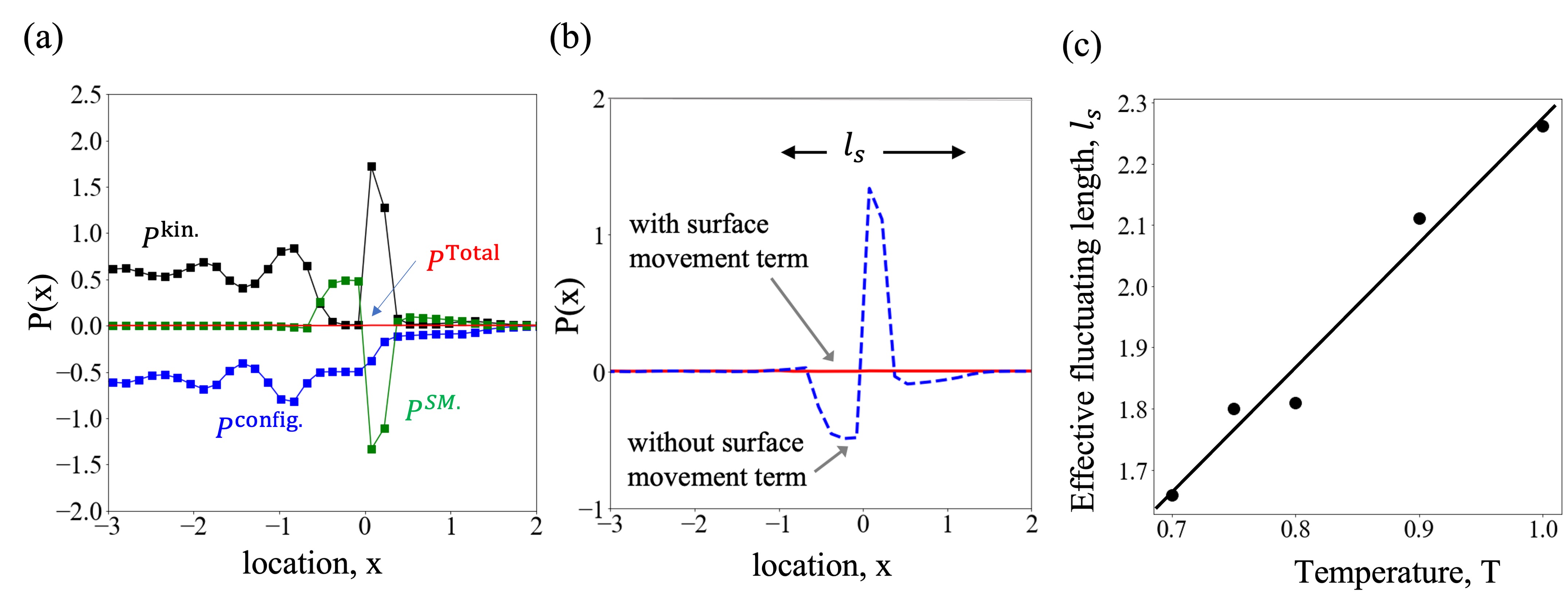}
    \caption{[Color online] Pressure profiles at and near the surface layer for T= 0.70. (a) Components of the normal pressure: (\protect\bluelineS\protect\squareB\protect\bluelineS) configurational part, $P^\mathrm{config.}$,   (\protect\blacklineS\protect\squareK\protect\blacklineS) kinetic part, $P^\mathrm{kin.}$, (\protect\greenlineS\protect\squareG\protect\greenlineS) contributions from surface movement, $P^\mathrm{SM.}$, and (\protect \redline) the total normal pressure, $P^\mathrm{Total}$. (b) Normal pressure without (\protect\bluelineDashed) and with (\protect\redline) the consideration of surface movement term, $l_s$ denote the effective fluctuating length, over which the surface movements are visible. (c) Linear increase of the effective fluctuating length, $l_s$ with temperature, the solid line is a linear fit to the data with a slope of $2.04$, an intercept of $0.23$, and $R^2= 0.97$)}
    \label{fig:SurfMov}
\end{figure}

The oscillatory nature of the intrinsic profiles can be further realized from the pressure profiles.
Both the normal and the tangential components of the pressure tensor can be decomposed into their constituent parts.
Fig.~\ref{fig:SurfMov}\,(a) shows these various components (averaged over time and $yz$ plane) of normal pressure for $T = 0.70$. 
To track the movement of the interface and its temporal evolution at a resolution of atomic spacing, ~\citet{smith2020hydrodynamics}, and \citet{smith2021importance} discarded the otherwise applied concept of an average interaction contour and introduced an instantaneous frame of reference which evolves with the interface and hence describes the pressure tensor in a purely mechanical manner. 
The consideration of the surface movement contributes to an additional corrective term, $P^\mathrm{SM.}$.
This, when considered along with the configurational and the kinetic contributions of the normal pressure, as shown in Fig.~\ref{fig:SurfMov}\,(a), exactly balances the momentum change to machine precision.

The essence of the surface movement contribution is illustrated in  Fig.~\ref{fig:SurfMov}\,(b) where the normal pressure is plotted without and with the consideration of the pressure correction. 
Only when the surface movement effects are accounted for, the kinetic and the configurational components can precisely balance each other resulting in a perfectly flat profile for the normal pressure which signifies that the liquid-vapor interface is mechanically stable.
It is apparent from Fig.~\ref{fig:SurfMov}\,(b) that the pressure contribution due to surface movements fluctuates only over a length of few atomic diameters (where the dashed-line shows fluctuations), and remains constant elsewhere.
We denote this length as the {\textit{effective fluctuating length}}, $l_s$ (schematically shown in Fig.~\ref{fig:SurfMov}\,b). 
More formally, we define $l_{s}$ as
\begin{equation}\label{eq:eff_fluc_len}
l_{s}(\epsilon)=\sup_{\mathbb{R}}\Bigl\{\left|x_{i}-x_{j}\right|:x_{i},x_{j}\in\{x:\left|P(x)\right|=\epsilon\}\Bigr\},
\end{equation}
which is the maximal distance between any two roots of the equation $\left|P(x)\right|=\epsilon$, for $0<\epsilon\ll1$ is the amplitude of the very small, but finite oscillations either side of the effective fluctuating length. In this study, we use a cut-off value $\epsilon=0.05$ in order to numerically determine $l_{s}$ from the fluctuations of $P(x)$.
As shown in Fig.~\ref{fig:SurfMov}\,(c), $l_s$ increases linearly with the temperature. 
Despite that the interfacial thickness assumes approximately the size of an atom at a temperature away from the critical temperature~\cite{lu2008molecular}, the effective fluctuating length demonstrates that the kinetic contribution spans over at least a few atomic diameters.
This is reasonably justified in light of the fact that a higher temperature corresponds to greater surface movements and thereby, a (kinetically) thicker interface~\cite{goujon2016can}.

Although not straight forward, a plausible link between the effective length, $l_s$ and the shear viscosity can be inferred through the hydrodynamic description of the capillary waves. It is well established~\cite{levich1962physicochemical, delgado2008hydrodynamics, shen2018capillary} that the damping rate, $\Gamma$ of an over-damped capillary wave mode satisfies the asymptotic relation, $\Gamma \sim q \gamma / \mu$ where $q$ is the wavenumber and $\mu$ is the dynamic shear viscosity. At critical damping, the real part of the complex wave frequency is zero, i.e., $\mathrm{Re} (\omega) =0$ and $q \sim \gamma \rho / \mu^2$. If the propagation velocity of the density fluctuations, $v_0$, is further assumed to be constant, it can then be postulated that $l_s$ scales inversely with $\Gamma$, i.e,~$\Gamma \sim {v_0/l_s} \sim \gamma^2 \rho / \mu^3$. Crucially, this rearranges to give 
    
    \begin{equation}
        l_s \sim \frac{v_0 \mu^3}{\gamma^2 \rho},\label{eq:l_s-mu}
    \end{equation}
thereby provides a link between the dynamic shear viscosity and the effective fluctuating length. This is an interesting interpretation of the effect of viscosity on the fluctuations of the interface at the atomic scale. However, a detailed analysis to confirm this postulate is beyond the scope of this manuscript but this relationship between $l_s$ and $\mu$ certainly prompts further investigation. 

\begin{figure}
    \centering
    \includegraphics[width=0.95\linewidth]{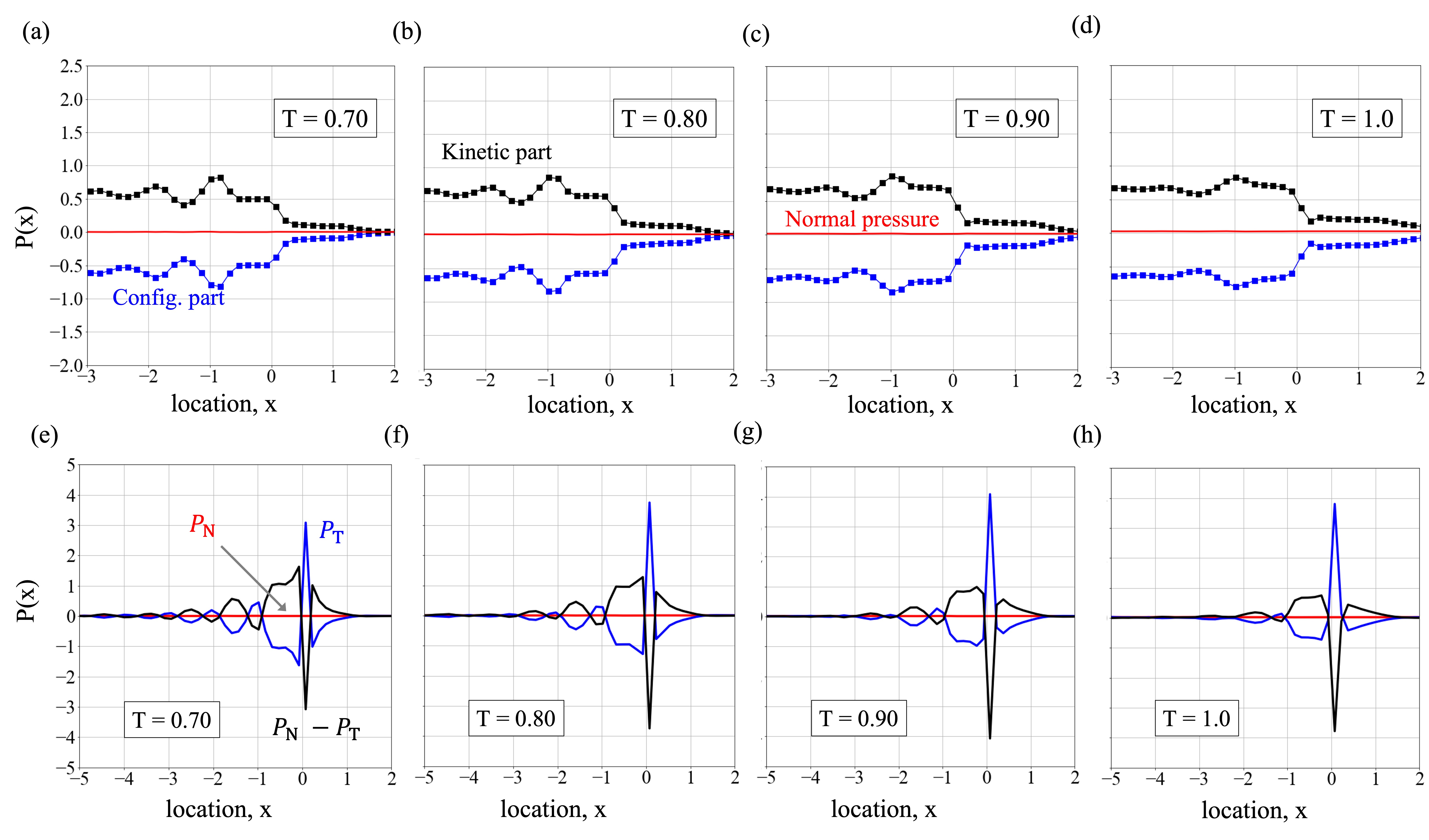}
    \caption{[Color online] (a-d) Constituent parts of the (\protect \redline) normal component of pressure, i.e.,  (\protect\blacklineS\protect\squareK\protect\blacklineS) the kinetic part and, (\protect\bluelineS\protect\squareB\protect\bluelineS) the configurational part for (a) T= 0.70, (b) T=  0.80,  (c) T=  0.90  and,  (d) T=  1.0. (e-h) Variation of (\protect\blueline) the tangential component of the pressure tensor and of (\protect\blackline) $P_N - P_T$  near the interface for $T = 0.70$ to $1.0$.}
    \label{fig:NormTan}
\end{figure}

The temperature effect on the intrinsic pressure profile has further been examined in Fig.~\ref{fig:NormTan}.
Though the shape of the normal pressure profile remains unaltered irrespective of temperature,  the oscillations of the configurational and the kinetic parts, in Fig.\,\ref{fig:NormTan} (a-d), are seen to smooth out as temperature increases.
The normal and the tangential components of the pressure, along with $(P_N-P_T)$, are plotted in the Fig.\,\ref{fig:NormTan} (e$-$h) for different temperatures. Interestingly, the tangential pressure profiles become less corrugated at higher temperatures, that is, the spatial oscillations or the expansion and contraction of the surface become less prominent.
Such behavior can be ascribed to the more energetic interface at higher temperature making the interface tracking difficult.
    The dyadic term in the kinetic pressure of Eq.\,(\ref{P_equation}) is a particle property, and thereby proportional to the intrinsic density through $P^\mathrm{kin.} = \rho k_B T$, where $k_B$ is the Boltzman constant. As such these oscillations are directly comparable to those in the intrinsic density profiles, as in Figs.\,\ref{fig:densityOsc} and \ref{fig:osc_fitting}. The equal and opposite of this kinetic component is the configurational pressure, as shown in Fig. 5 (a-d), which ensures equilibrium.
The observations made in Fig.~\ref{fig:SurfMov}\,(c) and Fig.\,\ref{fig:NormTan} allow to conclude that an increase in temperature dampens the oscillations of the intrinsic profiles, and at the same time, broadens the effective fluctuating length, $l_s$, over which surface movement effects are present.

\begin{figure}
    \centering
    \includegraphics[width=0.65\linewidth]{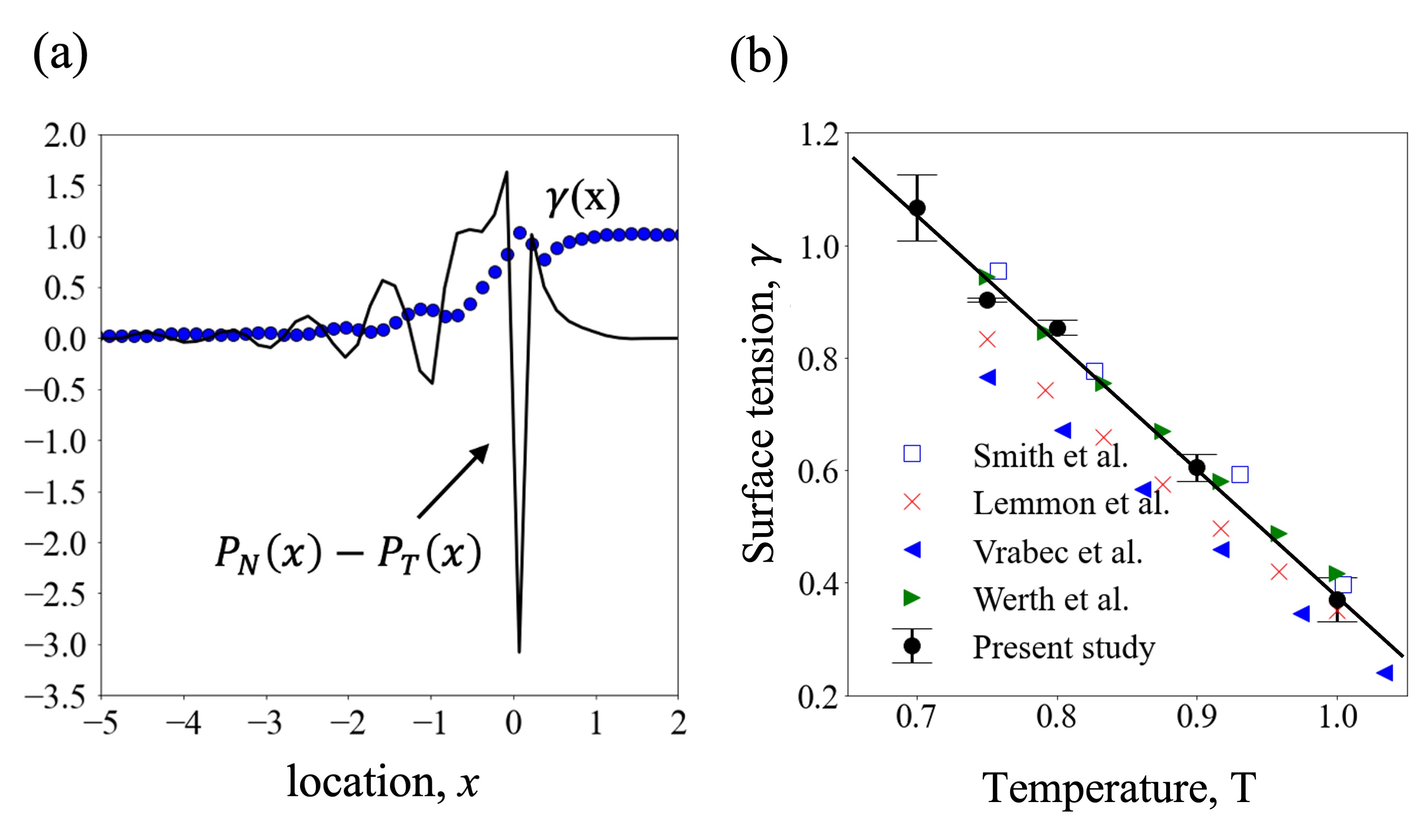}
    \caption{[Color online] (a) Difference between the normal and the tangential pressure,  $P_N - P_T$ (\protect\blackline) near the interface. The cumulative integral of $P_N - P_T$, i.e., the surface tension, $\gamma(x) = \int_{-10}^x \left[P_N (x^\prime) - P_T (x^\prime)\right] dx^\prime$ is shown by the filled circles (\protect\circleB)  where, the lower limit of integration is kept fixed at $x=-10$ and the upper limit is gradually increased up to $x=2$. (b) Surface tension from present modelling (\protect\circleK) shows good agreement with results from previous experiments and simulations. The error bars denote the standard deviation of surface tension as measured from three separate ensembles. The solid line is a linear fit (with slope $=-2.26$, intercept $=2.63$ and $R^2 = 0.99$) to the surface tension obtained from the present study.}
    \label{fig:cumul}
\end{figure}

The influence of the oscillations discussed, are confined to a small interfacial region.
Far from the liquid-vapor interface, both the normal and the tangential pressure become equal and uniform. Hence, it is not surprising that only the region of a few atomic diameters from the interface (both in the liquid and the vapor sides) contributes to the surface tension.
The term inside the integral of Eq.\,(\ref{Eq:surfTenMec}) is plotted (black solid line) in Fig.~\ref{fig:cumul}\,(a) along with the cumulative integral, i.e, the surface tension, $\gamma$.
The magnitude of the surface tension is seen to reach a plateau right after the peak, within an atomic diameter in the vapor-side. 
A careful examination of the intrinsic surface (i.e., $x = 0$) thus portrays its outright significance in determining the surface properties, as seen in Fig.~\ref{fig:cumul}\,(a), where the large negative peak of the pressure difference at $x = 0$ contributes significantly to the integral of Eq.\,(\ref{Eq:surfTenMec}).
We compare, in Fig.~\ref{fig:cumul}\,(b), the surface tension evaluated at different temperatures with available experimental data for liquid argon~\cite{lemmon1998thermophysical}, 
results from molecular simulations with truncated and shifted potential with a cut-off radius of 2.5~\cite{vrabec2006comprehensive,goujon2014gas}, and 4.0~\cite{smith2016langevin}, and truncated potential with long range corrections and cut-off radius of 3.0~\cite{werth2013influence}- as seen in the figure, the comparison suggests good agreement.

\subsection{Surface fractals}
\begin{figure}
    \centering
    \includegraphics[width=0.95\linewidth]{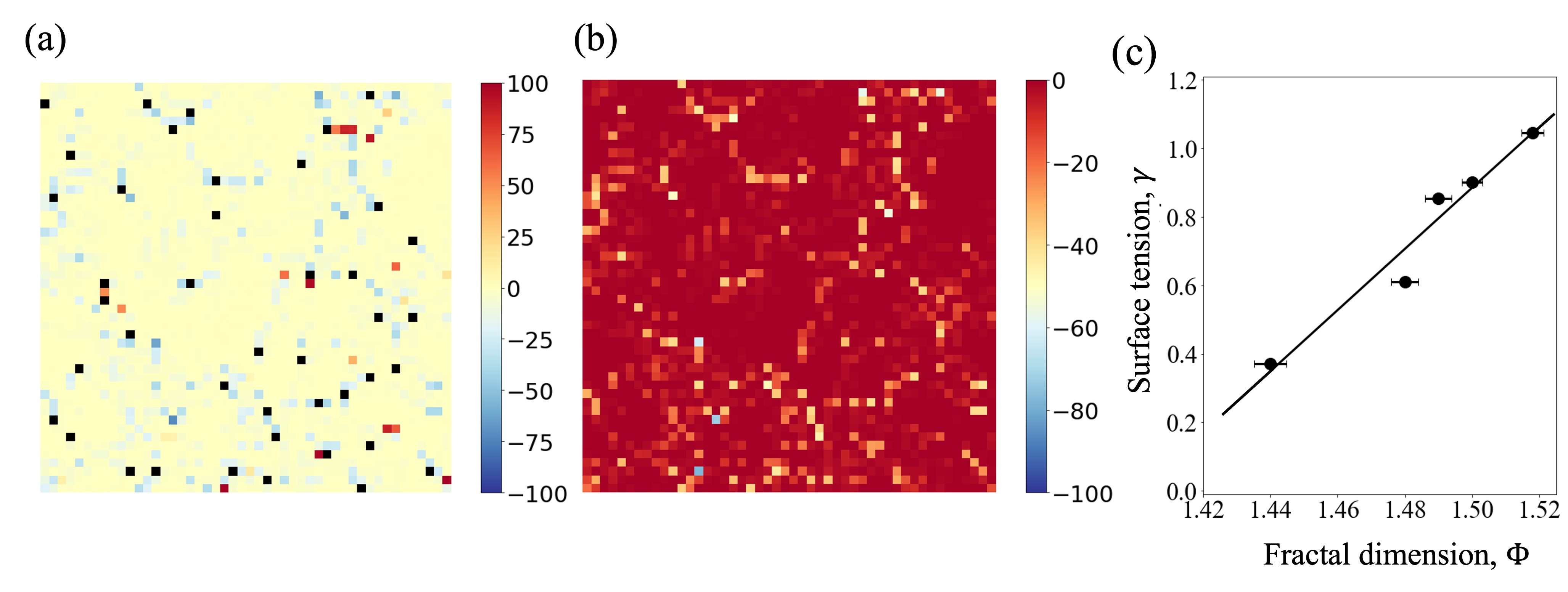}
    \caption{[Color online] Instantaneous stress map at the liquid-vapor interface for T = 0.80. (a) Color map of the configurational stress distribution with atomic positions overlaid (filled black squares). (b) Only negative configurational stresses, (c) surface tension as a function of the fractal dimension of stress network at the surface layer over the range of temperatures considered in this study, error bars (with a magnification factor of 2 for visualization purpose) denote standard error of the mean of fractal dimension, SEM, the solid line is a linear fit with $R^2 = 0.95$.}
    \label{fig:stressNetwork}
\end{figure}

The preceding sections discuss the temperature effects on the intrinsic profiles, and on the corresponding values of the surface tension. How such alterations take place, however, remain unanswered until now.
To investigate this, and because the configurational stresses are the sole contributors towards the surface tension, we examine the distribution of these stresses and ignore the kinetic components, at the intrinsic surface layer ($x=0$); i.e.,
\begin{displaymath}
   P^\mathrm{config.} (x=0) =  \frac{1}{2} \left[ P_\mathrm{yy}^\mathrm{config.} (x=0) + P_\mathrm{zz}^\mathrm{config.} (x=0) \right] .
\end{displaymath}
Fig.~\ref{fig:stressNetwork}(a) shows an instantaneous configurational stress distribution map at the interface. The filled black squares correspond to the atomic locations which the intrinsic interface is fitted to, at that particular instant. 
In Fig.~\ref{fig:stressNetwork}(b), only stresses that contribute to the tension of the surface (negative stresses) are shown, where a cluster spreading over the surface becomes apparent (also see Fig.\,S1, S2, movie\,S1 and related discussions).
These `non-trivial' networks and their variation with temperature are analysed by means of fractal analysis.  

To quantify the temperature dependency of the structural complexity of the stress networks, we measure their fractal dimensions, $\Phi$. This dimension is often used to quantify the space filling nature, heterogeneity or self-similarity of surfaces, clusters etc.~\cite{avnir1984molecular, jelinek2006understanding,chen2015fractal,shen2020transient}
$\Phi$ is calculated through the Minkowski dimension~\cite{hassan2012bioinspired,kempkes2019design}, which is often referred to as the box-counting dimension, whereby, for non-overlapping $N$ boxes with sides $\epsilon$,
\begin{equation}\label{eq:boxcount}
    \Phi = \lim_{\epsilon \to 0} \frac{\log N (\epsilon)}{\log (1/\epsilon)}.
\end{equation}
The algorithm employed to calculate $\Phi$ consists of converting the network maps (as in Fig.\,\ref{fig:stressNetwork}\,b) into binary images such that the stress networks are depicted by black pixels on white background. For a given grid side, $\epsilon$ the number of grids, $N(\epsilon)$ required to fill the projected surface area of the aggregate is counted and the grid is made increasingly finer at each subsequent iteration. The fractal dimension, $\Phi$ is then obtained from the slope of $\mbox{log}(N)$ vs $\mbox{log}(\epsilon)$. See  supporting information for further details.

With the variation of temperature, we have observed consistent behaviour of the fractal dimension of the network comprised of the negative configurational stresses that lie within $\left[ P^\mathrm{config.}_\mathrm{n,m} - \sigma,  P^\mathrm{config.}_\mathrm{n,m}\right]$, where $P^\mathrm{config.}_\mathrm{n,m}$ denotes the mean of all the negative configurational stresses, and $\sigma$ denotes the standard deviation.
The fractal dimensions thus obtained are compared against the corresponding surface tensions in
Fig.~\ref{fig:stressNetwork}\,(c). 
For the temperature range considered here, $\Phi$ is seen to linearly correspond to the surface tension as $\gamma = 9.86\, \Phi\ - 13.86$ which is shown by the solid line.
This reflects the greater space filling nature~\cite{jelinek2006understanding} of the (surface) stress network at a lower temperature. 
At the same time, the fractal dimension of the stress clusters only at the outermost atomic layer proves to be predictive of the surface tension.

In hydrodynamics, the surface tension, $\gamma$ is often modelled as an equation of state in terms of temperature, surfactant concentration etc. This seemingly mechanistic approach does not capture the subtleties of the mutual interactions between the various effects nor does it directly model the fundamental thermodynamic nature of surface tension, that is, the free energy it would cost to form an interface. On the other hand, a localised surface fractal stress approach to surface tension via the calculation of localised fractal dimensions of the interface stress distribution would improve on both of these issues.
Firstly, in a system where thermo- and soluto-Marangoni effects are in play, the localised fractal stress would take into account both of these two effects and their interactions with each other without any loss of generality. Secondly, the localised fractal stress can provide a standardised platform upon which we examine all possible effects on surface tension (whether local or global) in an uniform and consistent way, in this sense, a higher localised fractal stress can be directly interpreted as a higher energy cost required to form the interface in that region, thereby resulting in a higher localised surface tension. 
A complete hydrodynamical description of this interesting problem is not pursued in this study, but we anticipate numerous applications of this approach to the modelling of nano-fluidic interfacial phenomena with a highly variable localised surface tension which would be realised in a future contribution.

\begin{figure}
    \centering
    \includegraphics[width=0.95\linewidth]{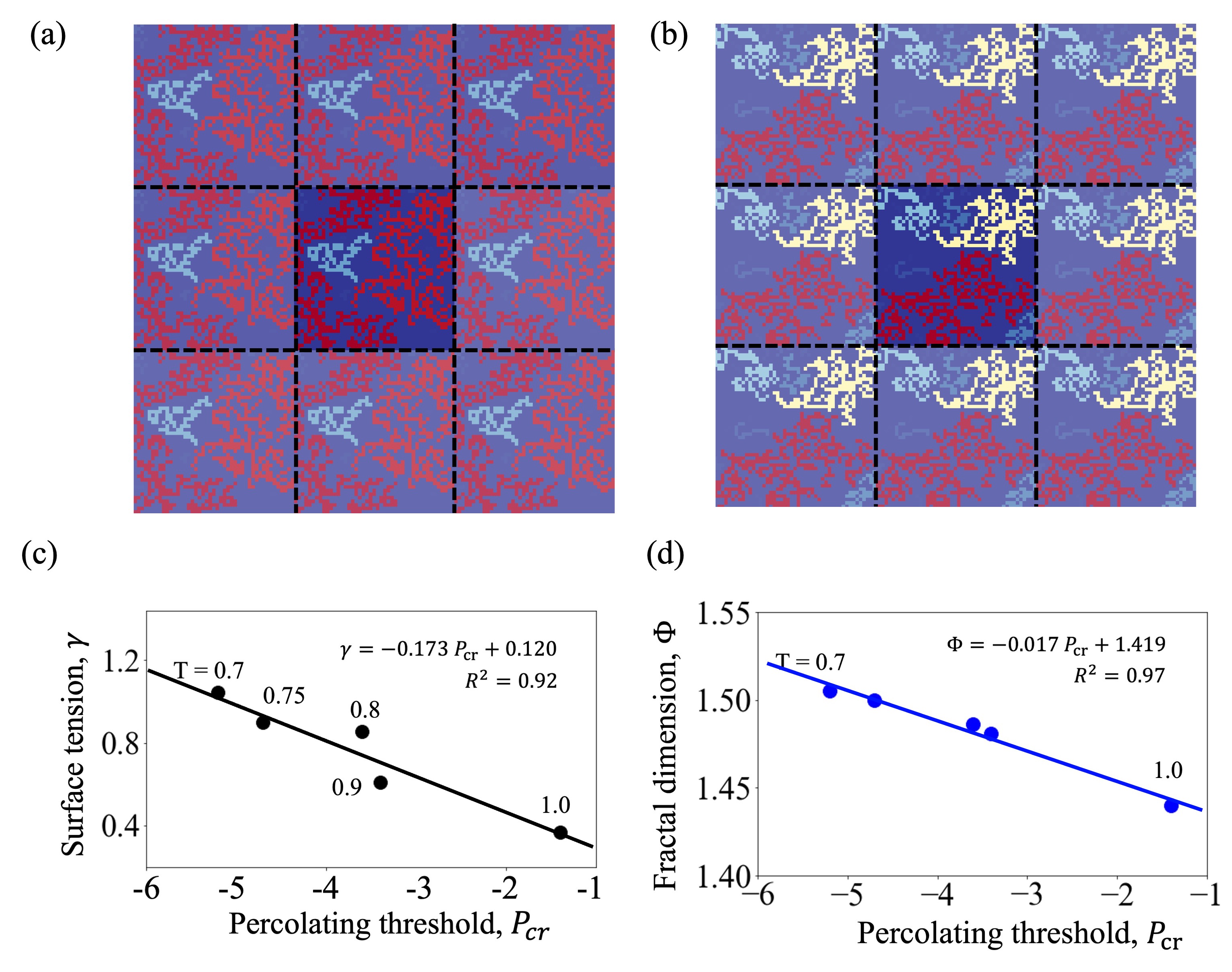}
    \caption{[Color online] Percolation network, with periodic images shown faintly, at a random instant at temperature (a) T = 0.8 and, (b) T = 1.0.  The largest cluster is colored red. Panel (a) shows a spanning cluster, whereas, in panel (b) no cluster is spanning throughout the surface for the same stress threshold.
    (c) Surface tension, and (d) fractal dimension as functions of the percolation threshold over the range of temperatures considered in this study, lines are linear fits.}
    \label{fig:perc}
\end{figure}

\subsection{Surface percolating clusters}
The accurate identification of the intrinsic surface layer allows us to further the analysis of the atomic interaction network by applying the concept of percolation~\cite{pathak2017force, sega2014two,stauffer2018introduction,heyes1988percolation,de2009rigidity,dapp2012self}. This method examines the `connectedness' of the different sites (or cells) on the interface that experience stress lower than a threshold.
For a critical stress, a connected network of sites or a spanning cluster is formed that spans from left to right and from top to bottom of the lattice. If such spanning clusters form in at least $50$ percent of the configurations, a system is described as a percolating system~\cite{zarragoicoechea2005percolation,seaton1987aggregation} and the corresponding stress is known as the percolating threshold~\cite{bug1985interactions,heyes1988percolation,heyes1989microscopic,heyes1990continuum,hasmy2021percolation}. 
Here, by configuration, we mean the instantaneous behaviour of the system captured at a single time step.
We determine the stress percolating threshold at the intrinsic interface and assess how temperature effects this threshold.

To quantify the threshold we consider percolation in both directions (left-to-right and top-to-bottom) and
apply next-nearest neighbours algorithm. Fig.~\ref{fig:perc} (a) and (b) show two randomly selected instances, respectively for $T = 0.8$ and $1.0$, where spanning clusters form for the case of $T = 0.8$, but no such cluster is seen for $T = 1.0$.
Such networks can be interpreted as a manifestation of the stress heterogeneity at the surface, which essentially increases with temperature resulting in a lower surface tension.
Indeed, at lower temperature, the majority of the surface experiences higher negative stress, and thereby a spanning network can form at a relatively larger (negative) stress threshold.
On the contrary, as temperature increases, the heterogeneity in stress distribution increases too, and hence the formation of a spanning network requires the inclusion of smaller clusters. 
Fig.~\ref{fig:perc} (c) illustrates this, where the surface tension is plotted as a (linear) function of the percolating threshold. A lower value of the surface tension (which corresponds to a higher temperature) is seen to be associated with a lower (negative) percolating threshold, whereas, a higher surface tension (lower temperature) is related to a larger magnitude of the threshold.
Similar functional dependency between the percolating threshold and the fractal dimension of the stress networks can be seen in Fig.~\ref{fig:perc}\,(d), which further echos the higher space filling nature of the surface clusters with lower stress heterogeneity at a lower temperature. 

\begin{figure}
    \centering
    \includegraphics[width=0.90\linewidth]{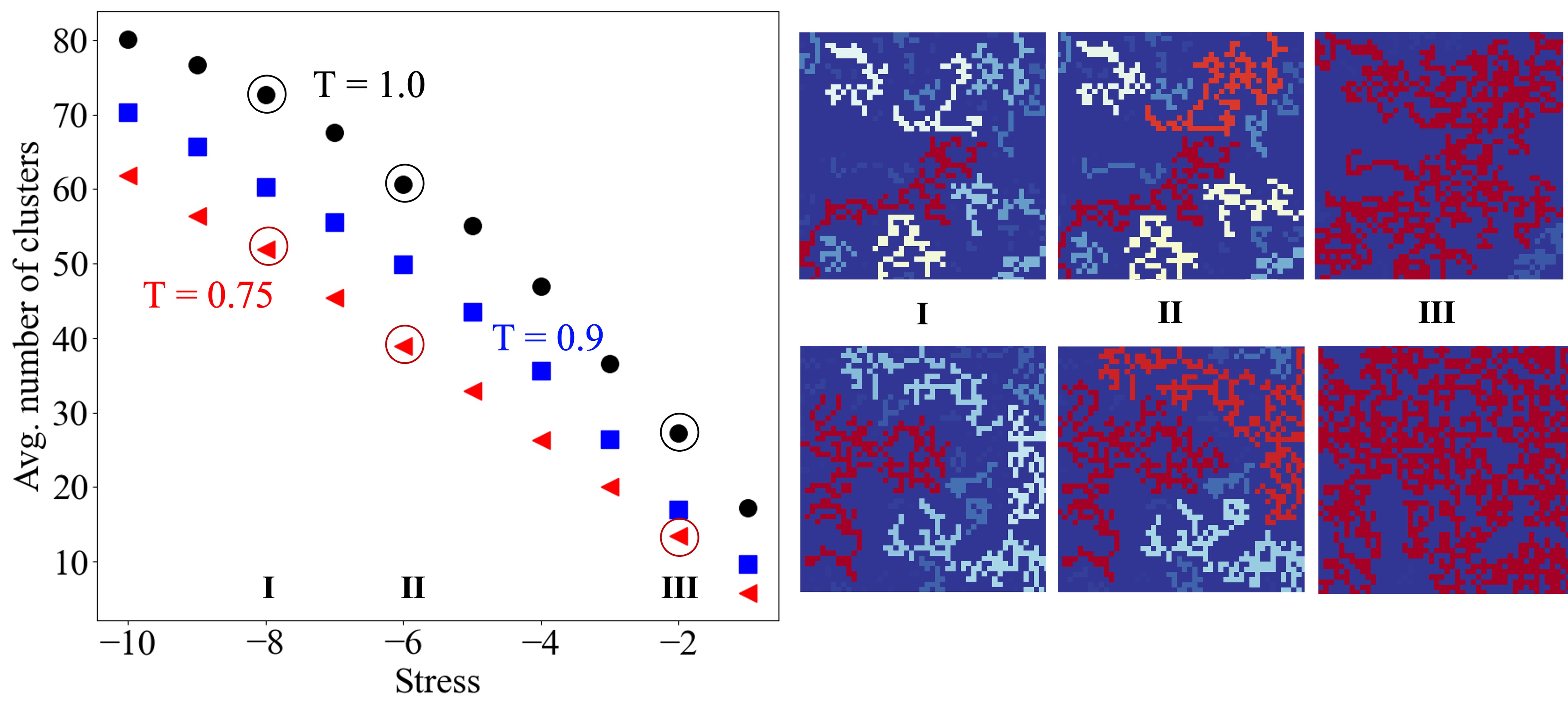}
    \caption{[Color online] Average number of clusters at different stresses for $T = 0.75$ (red triangles), $0.9$ (blue squares) and $1.0$ (black circles). The top panel in the right hand side corresponds to three instantaneous networks formed with stresses $\mathrm{I:} (-\infty, -8]$, $\mathrm{II:}(-\infty, -6]$ and $\mathrm{III:}(-\infty, -2]$ for $T = 1.0$. The bottom panel shows similar networks but for $T = 0.75$. The average number of clusters associated with I, II and III are circled in the left panel.} 
    \label{fig:cluster}
\end{figure}

Not only the stress networks at distinct temperatures differ at the percolating threshold, but their ability of cluster formation also varies at any stress.
In Fig.~\ref{fig:cluster}, the average number of clusters are plotted for a range of stresses. It is evident that for any particular stress, the (average) number of clusters at a lower temperature is less than that at a higher temperature, meaning that the low temperature clusters are larger and less heterogeneous, in terms of stress, than their high temperature counterparts.
For instance, the top panel in the right hand side of Fig.~\ref{fig:cluster} shows three instantaneous networks for (negative) stresses with magnitudes greater than $8, 6$ and $2$, with the largest clusters colored in red. 
It is seen that the clusters grow in size as the field becomes more inclusive of stresses from panel $\mathrm{I}$ to $\mathrm{III}$. Similar growth of the cluster size is seen for a lower temperature surface, i.e., for $T=0.75$ as in the bottom panel. However, what differs between the networks at the two temperatures of interest here is: for identical range of stresses, it is more probable to obtain a larger cluster for a lower temperature surface, see Fig.\,S3 and related discussions.

From a thermodynamic point of view, a liquid-vapor interface at a higher temperature is (thermally) more energetic\cite{palmer1976effect} and thereby, favours a broader interface and weaker spatial correlation between atoms with larger nearest-neighbours distances causing a lower surface tension~\cite{bhanvase2021101}. 
This, however, is challenging to perceive via a mechanical route - the inherent difficulty lying in separating the actual surface layer from the capillary fluctuations~\cite{sega2014two}.
Such complication is circumvented in this study by detecting the interface using the intrinsic sampling method and collecting stresses in a reference frame moving with the surface which establishes the ground for an authentic surface-network analysis.
Through this route, a thermodynamically weaker surface (due to the weaker spatial correlation between atoms) is mechanically represented, by the lower  `connectivity'~\cite{burnley2013importance,alvarado2017force} or higher heterogeneity of the stress networks. Stated differently, a high temperature surface (of which, surface tension is lower) can be mechanically described as a surface that is loosely inter-connected by disjointed networks of stresses.
Although the sensitivity of the analysis is limited by the probabilistic nature of the quantities investigated, the sole analysis of the sharpest atomic description of the intrinsic interface proves to be useful to unravel the variation of surface properties with temperature.

Whereas the intriguing notion of interpreting surface properties through surface coverage remains challenging~\cite{mu2006neutron,menger2011relationship}, the stress network approach presented in this paper is quite straight forward. Simplification of the interface by modeling an LJ fluid is a key limitation of this study, but the approaches developed here could be extended to molecular fluids in future work.

\section{Conclusion}\label{sec:conclusion}

From the results of MD simulations of the liquid-vapor interface of an LJ fluid, we present a mechanical interpretation of surface tension and its variation with temperature.
Intrinsic sampling method is used to define a moving frame of reference and an equation for the interfacial density, $\rho = \rho(x,T)$ for $T\in [0.7,1.0]$ and $x\in(-\infty,-1]$ is presented.
We have identified stress-clusters at the intrinsic surface, and analyzed their non-uniform spatial correlations.
The atomic interactions at the intrinsic surface layer can be thought of as a network of stresses holding the surface altogether. 
At an elevated temperature, the network becomes more disjointed and thus less stable.

Both the fractal dimension and the percolating threshold of the stress network can be correlated with the surface tension. These observations suggest that the pattern formation and the connectivity of the stress networks at the intrinsic interface are good indicators of surface tension. Importantly, the analysis of the stress-network only at the outermost atomic layer suffices to provide a consistent prediction over a range of temperatures.
The surface tension acquired from the simulation agrees well with previous MD simulations using different methods and experimental data for liquid argon, which advocates for the extrapolation of the findings of this study to liquid-vapor interfaces of molecular systems of interest.

\section{Conflict of Interest}
The authors declare no conflict of interest.

\subsection{Supporting Information} 
\begin{itemize}
    \item[--] Fractal Analysis, Percolation Network, Liquid-Vapor coexistence data (PDF).
    \item[--] Temporal variation of fractal networks (Movie).
\end{itemize}

\subsection*{Acknowledgement}
The authors wish to thank Prof. D.M. Heyes (Department of Mechanical Engineering, Imperial College London, UK) for insightful discussions.
M.R.R. acknowledges PhD studentship funding from Shell via the University Technology Centre for Fuels and Lubricants and the Beit Trust for the Beit Fellowship for Scientific Research. L.S. thanks the Engineering and Physical Sciences Research Council (EPSRC) for a Postdoctoral Fellowship (EP/V005073/1). J.P.E. was supported by the Royal Academy of Engineering through a Research Fellowship. D.D. thanks the EPSRC for an Established Career Fellowship (EP/N025954/1).

\bibliography{RefLib}

\end{document}